\documentclass[twoside]{LCWS11}
\usepackage[latin1]{inputenc}
\usepackage{wrapfig,rotating}
\usepackage{amssymb,amsmath,array}
\usepackage{cite}
\usepackage{graphicx}
\usepackage{subfigure}

\begin{document}
\title{Recent results of Micromegas sDHCAL\\
with a new readout chip}

\author{C.\,Adloff$^1$, J.\,Blaha$^1$, J.-J.\,Blaising$^1$,  M.\,Chefdeville$^1\footnote{Corresponding author: chefdevi@lapp.in2p3.fr}$, A.\,Dalmaz$^1$, C.\,Drancourt$^1$,\\
A.\,Espargili\`ere$^1$, R.\,Gaglione$^1$, N.\,Geffroy$^1$, J.\,Jacquemier$^1$, Y.\,Karyotakis$^1$, F.\,Peltier$^1$,\\
J.\,Prast$^1$, S.\,Tsigaridas$^2$, Y.\,Tsipolitis$^2$ and G.\,Vouters$^1$
\vspace{.3cm}\\
1- Laboratoire d'Annecy-le-Vieux de Physique des Particules - CNRS/IN2P3
\vspace{.1cm}\\
2- National Technical University of Athens}
\maketitle

\begin{abstract}
Calorimetry at future linear colliders could be based on a particle flow approach where granularity is the key to high jet energy resolution.
Among different technologies, Micromegas chambers with 1\,cm$^{2}$ pad segmentation are studied for the active medium of a hadronic calorimeter.
A chamber of 1\,m$^{2}$ with 9216 channels read out by a low noise front-end ASIC called MICROROC has recently been constructed and tested.
Chamber design, ASIC circuitry and preliminary test beam results are reported.

\end{abstract}

\section{Introduction}

\subsection{Calorimetry at a future linear collider}

The detailed study of electroweak symmetry breaking and of the properties of a hypothetical SM Higgs boson are some of the physics goals motivating the construction of a linear electron collider.
The centre-of-mass energy of the collider would be 500\,GeV (ILC) or 3\,TeV (CLIC), which will be decided upon analysis of the LHC data \cite{ILC_RDR,CLIC_CDR}.
In both cases, however, several interesting physics channels will appear in multi-jet final states, often accompanied by charged leptons and missing transverse energy.
The di-jet energy resolution should be such that \textit{Z} and \textit{W} masses can be determined with an accuracy comparable to the natural decay width of these bosons.
This translates into a jet energy resolution of 3--4\,\% over the whole energy range.
Two kind of calorimeters are under study to meet this requirement \cite{CALO}.
One proposes to measure and correct for the fluctuations of the electromagnetic and hadronic components on a shower by shower basis.
The second is based on the \textit{Particle Flow} (\textit{PF}) approach: the momentum of jet charged particles are measured with the tracking system, tracks and showers being matched thanks to a high segmentation of the calorimeters.

\subsection{\textit{Particle Flow} calorimeters}

Several technological options are studied by the CALICE collaboration: silicon or scintillator for a sampling ECAL with tungsten absorbers, scintillator or gaseous chambers for the Fe or W HCAL. In the gaseous detector case, R\&D on RPC, GEM and Micromegas are actively pursued.
They are an attractive option with respect to the large area to be instrumented: 3000\,m$^{2}$ in the SiD detector concept \cite{SID}.
In addition, because there is only one sensor per layer (a gas layer), higher cell segmentation can be achieved than with scintillators (1 versus 9\,cm$^{2}$) with plausibly better \textit{PF} performance.
On the other hand, due to the large signal fluctuations in gaseous detectors, the charge information is of limited use.
This may limit the single particle energy resolution necessary for measuring the neutral content of jets.
As a result, gaseous hadronic calorimeters are essentially digital (1--2\,bits per channel).

\subsection{Micromegas for a multi-threshold HCAL}

In the last ten years, the so-called Bulk fabrication process has widened the field of application of Micromegas detectors \cite{BULK}.
It consists in the lamination of a stainless steel mesh (20\,$\mu$m diameter wires woven to a square pattern) and photosensitive insulating films onto a PCB.
After patterning of the films to pillars, the mesh is strongly encapsulated in between the pillars at a constant distance from the anode PCB plane.
It is therefore well suited for manufacturing large area Micromegas
detectors. Furthermore, the final assembly is robust, the mesh
resistant to sparks and the gain uniformity relatively good, about $\sim$\,10\,\% \cite{ESPAR}.

\vspace{0.5mm}

Attractive features of Micromegas for a hadron calorimeter are excellent performance to MIP, high rate capability and, in suitable gas mixtures, no sentivity to neutrons and negligible aging.
As mentioned already, the energy measurement is relevant for neutrals but also for very high energy charged particles against which the tracker resolution worsen.
There, the energy is measured by calorimetry only and it is desirable to minimize both stochastic and constant terms (\textit{a}/$\sqrt{E}$\,$\oplus$\,\textit{b}) that govern the resolution.
In order to improve \textit{a} and linearity, it has been proposed to digitize signals with 2\,bit resolution.
This semi-digital readout scheme has been adopted by Micromegas and RPC HCAL projects (sDHCAL).

\section{The one square meter Micromegas prototype}

\subsection{Active sensor units}

\begin{figure}[h!]
\begin{centering}
\includegraphics[width=0.38\textwidth]{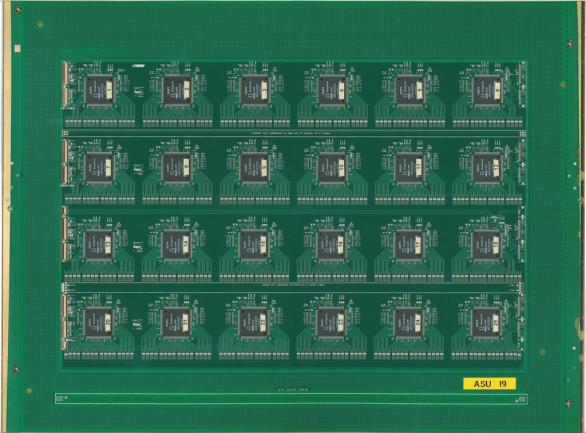}
\includegraphics[width=0.38\textwidth]{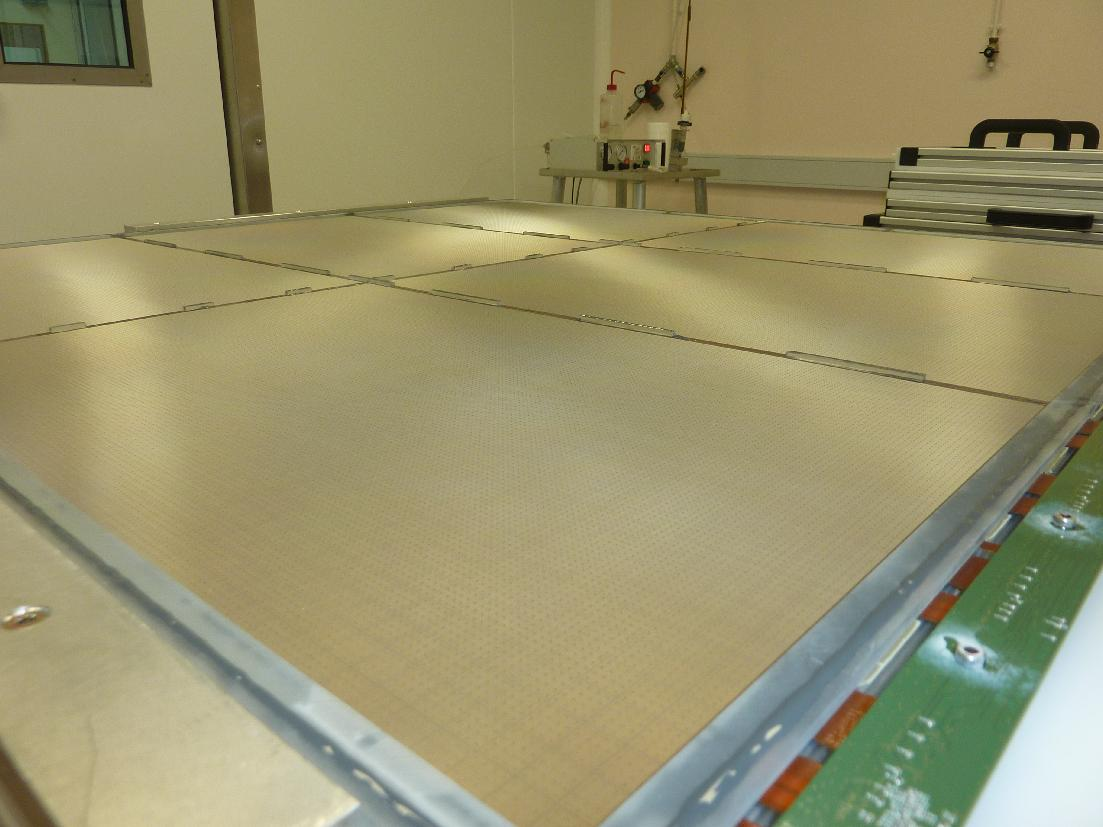}
\caption{Photographs of an Active Sensor Unit and the 1\,m$^{2}$ prototype during assembly.}
\label{m2_pics}
\end{centering}
\end{figure}

Lamination of the mesh and the PCB between rolls imply that the Bulk process is applied to flat PCB.
Hermeticity requirements and huge number of channels of an ILC/CLIC HCAL, however, make it unavoidable to integrate front-end ASIC on detectors.
Therefore, LAPP group and CERN workshop made the Bulk process compatible with ASIC equipped PCB.
The final assembly, called an Active Sensor Unit or ASU, consists of a PCB with 1\,cm$^{2}$ pads plus mesh on one side and ASICs on the other side (Fig.\,\ref{m2_pics} (left)).

The first ASU was constructed in 2008. It consisted of a matrix of 8\,$\times$\,8 pads read out by one DIRAC
chip and was successfully tested in CERN/PS particle beams \cite{DIRAC,HARDROC}.
Subsequently, ASUs of larger area were fabricated and equipped with HARDROC2 chips \cite{TIPP}. The size adopted for the construction of a 1\,m$^{2}$ detectors is 32\,$\times$\,48\,cm$^{2}$. Optimized for the efficient detection of Micromegas signals, a new ASIC called MICROROC was developed by the LAL/Omega group and the LAPP electronics department.

\subsection{Mechanical design}

The one square meter prototype features 9216 readout channels (96\,$\times$\,96 pads). It is assembled from six ASU glued on a stainless steel supporting plate and surrounded by a plastic frame (Figure~\ref{m2_pics} (right)). Another stainless steel plate holds a copper drift electrode. The 3\,mm drift gap is defined by the frame height and is kept constant over all chamber area by tiny spacers placed between the ASU. Each ASU consists of woven mesh and a PCB with 48\,$\times$\,32 readout pads of 1\,$\times$\,1\,cm$^2$ and 24 MICROROC chips.

\vspace{1mm}

The readout of two ASU is chained serially and connected to the data
acquisition system by three detector interface boards (DIF). The DIF
is a mezzanine board, it allows to load the ASIC configuration,
readout the data from the ASIC memory and also to provide system clock
and power. Another board, called inter-DIF, is placed between the DIF
and the ASU to provide voltage to drift and mesh electrodes
($\leq$\,500\,V). The gas is distributed by one inlet and outlet traversing the frame of the chamber whose total thickness is about 1\,cm.

\section{Circuitry of the MICROROC ASIC}

\begin{figure}[htb]
  \centering
  \includegraphics[width=0.65\columnwidth]{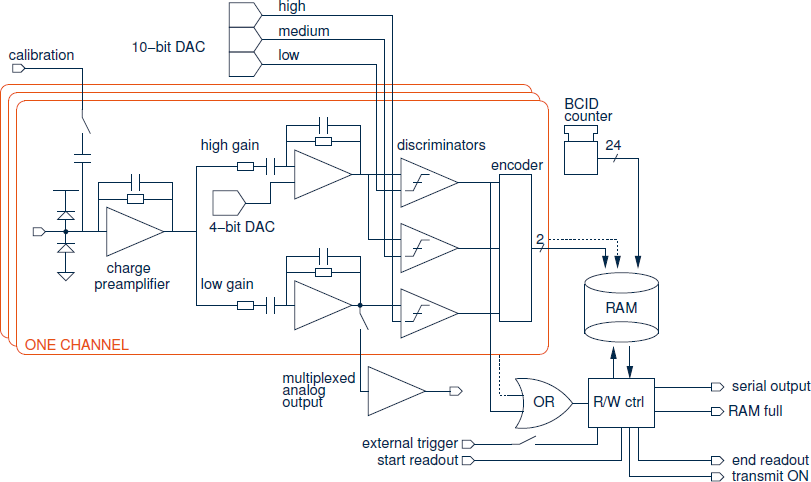}
  \caption{Schema of the MICROROC circuitry.}
  \label{mr_circuitry}
\end{figure}

The prototype is equipped with a newly developed 64-channel ASIC called MICROROC. Each ASIC channel features a diode network embedded inside the silicon to protect against discharges, a charge preamplifier followed by 2 shapers of different gain and shaping time tunable between 75 and 200\,ns, 3 discriminators allowing setting of 3 readout thresholds with 10-bit DACs and a 127 event depth memory with 200\,ns time-stamping of hits (Fig.\,\ref{mr_circuitry}). The dynamic range of the low and high gain shaper are 200 and 400\,fC respectively.

The output of the high gain shaper is connected to 2 discriminators which are used to define the low and medium channel threshold while a third discriminator, linked to the low gain shaper, is available to set the high threshold.
The 3 thresholds are common to the 64 chip channels. A 4-bit offset, however, can be used to vary the individual pedestal positions. It virtually provides a channel to channel control of the 3 thresholds.
A detailed characterization of 341 ASICs was carried out to verify
their overall functionality and determine their settings (offsets and thresholds).
The procedure and the results can be found in \cite{TIPP}.

\newpage

\section{Test of the prototype in particle beams}

\subsection{Experimental conditions}
\label{setup_h4}

A telescope and the 1\,m$^{2}$ prototype were installed in SPS/H4 for 18 days during August 2011.
At an average trigger rate of 100\,Hz in spill, 6 millions muons and pions in the ratio 85/15 were recorded.
Muons were used to assess the prototype performance to MIPs under various detector settings.
Shower signals were measured with a pion beam focused at a small iron block placed 0.5\,m upstream of the prototype.

\vspace{1mm}

\begin{figure}[h!]
\begin{centering}
\includegraphics[width=0.35\textwidth]{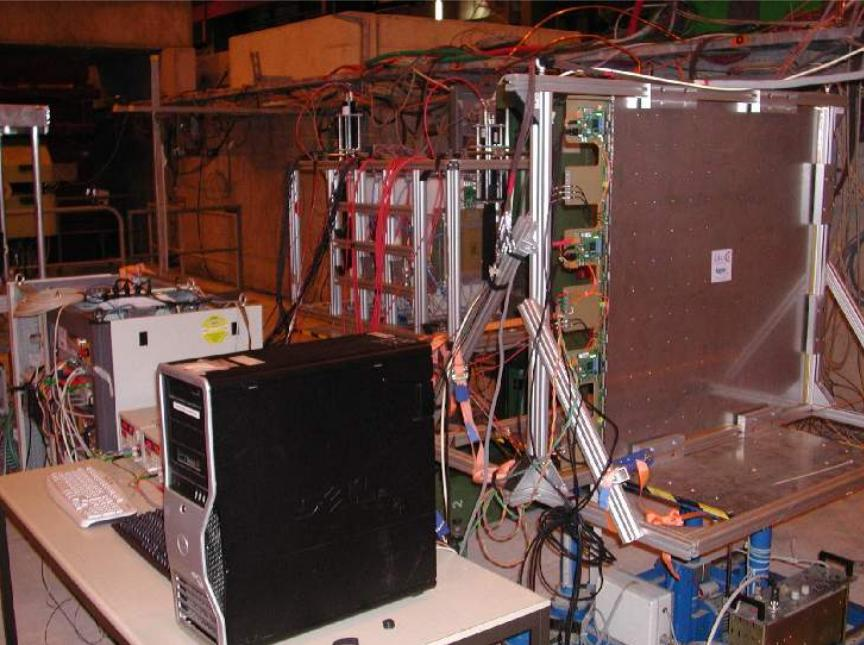}
\includegraphics[width=0.42\textwidth]{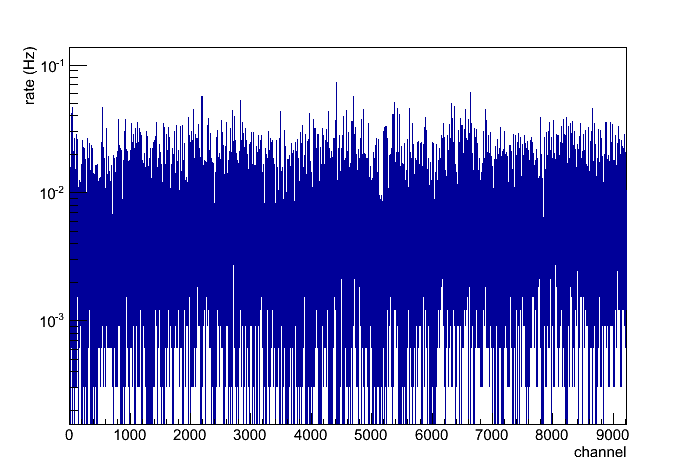}
\caption{Photographs of the summer 2011 setup in SPS/H4 (left). Noise rate over all channel prototype after pedestal alignment (right).}
\label{setup_noise}
\end{centering}
\end{figure}

Installed in the SPS/H4 beam line, the set-up consisted of a telescope and the prototype each placed on a movable table, a gas distribution panel, two racks with trigger electronics and power supplies and a fast acquisition PC (Figure~\ref{setup_noise} left). The telescope is a mechanical structure holding small Micromegas tracking chambers equipped with pads or strips for tracking and three scintillators plus PMT for triggering. All the detectors were flushed with a non flammable gas mixture of Ar/CF$_{4}$/\textit{i}C$_{4}$H$_{10}$ 95/3/2 at a total flow of 4\,l/h.

The iterative procedure to align the pedestals described in \cite{TIPP} was run at the beginning of the test (Fig.\,\ref{setup_noise} (right)).
Later, noise conditions were monitored with daily short dedicated runs.
The target noise rate was 10\,mHz\,/\,channel. Compared to a particle rate of 10\,Hz per channel in the center of the muon beam, it yields a high signal to noise ratio of about 10$^{3}$.

\subsection{Performance to MIPs}

\begin{figure}[h!]
\begin{centering}
\includegraphics[width=0.45\textwidth]{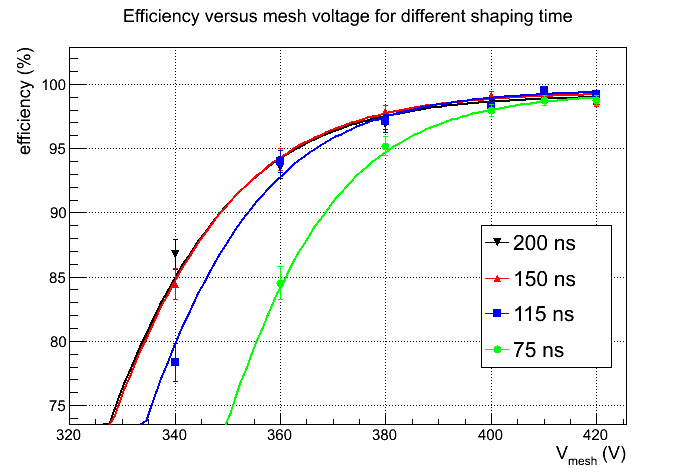}
\includegraphics[width=0.45\textwidth]{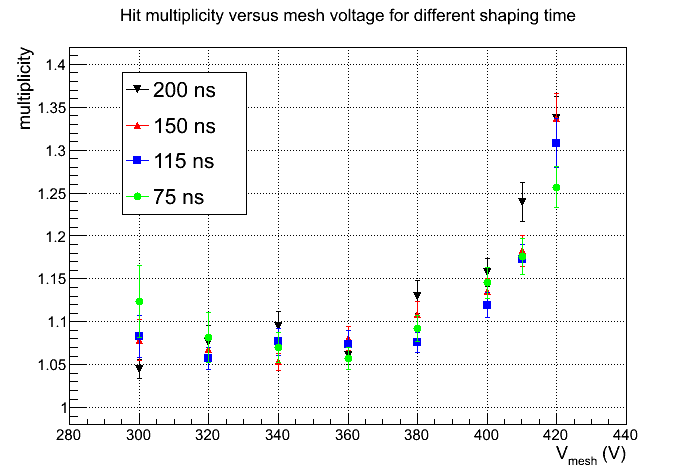}
\caption{Efficiency and pad multiplicity to muons at various operating conditions.}
\label{eff_mult_vmesh_scan}
\end{centering}
\end{figure}

\paragraph{Mesh voltage scan}
The efficiency is measured by finding a muon track in the telescope, extrapolating its impact point at the prototype and searching for hits inside a 3\,$\times$\,3 pad area at the time of the trigger (window of 600\,ns). The multiplicity is calculated as the average number of hits whenever at least one hit is present. Both quantities were measured at various mesh voltages and shaping times with the beam spreading over roughly hundred pads. A preliminary analysis leads to the trends showed in Figure~\ref{eff_mult_vmesh_scan} where a high efficiency is reached at all shaping settings. At 390\,V mesh voltage (gas gain around 3000), 1\,fC hit threshold and 200\,ns shaping, one records an efficiency of 98\,\% and a multiplicity of 1.12. Both are very well compatible with small prototype performance previously measured in similar conditions \cite{ESPAR}.

\begin{figure}[h!]
\begin{centering}
\includegraphics[width=0.33\textwidth]{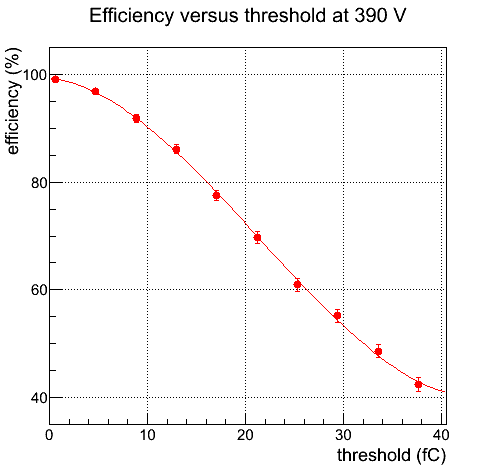}
\includegraphics[width=0.33\textwidth]{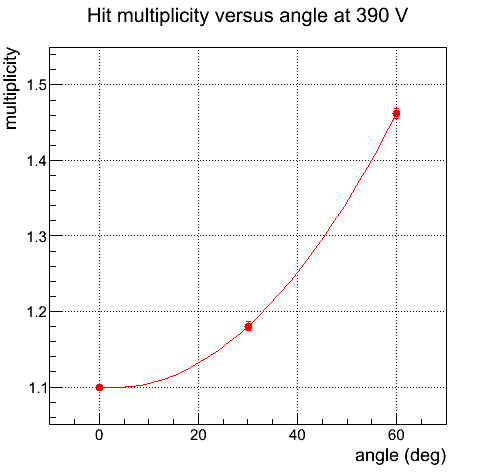}
\caption{Efficiency and multiplicity: threshold and angular scans.}
\label{eff_thr}
\end{centering}
\end{figure}

\paragraph{Threshold and angular scan}
Thanks to a careful calibration of the electronics before the test beam, a threshold scan could be performed (Figure~\ref{eff_thr} (left)).
The slight inflexion point at half of the range follows from the Landau distribution turnover.
The lowest running threshold is about 1\,fC.
Further increase to 2\,fC leads to a dramatic drop of the noise rate by 3 orders of magnitude while leaving the efficiency almost unchanged.
\noindent
Pad multiplicity is directly impacted by the angle of incidence of traversing muons.
Its dependence to the angle was studied by rotating the prototype with respect to the beam direction.
As is seen in Figure~\ref{eff_thr} (right), it remains below 1.5 at angles as large as 60$^{\rm \circ}$.

\subsection{Performance to pions}

Directed at a 20\,cm thick block of iron downstream of the telescope, a 150\,GeV/c pion beam was used to produce hadron showers which would propagate to the prototype placed half a meter downstream. This set-up allows to get a first idea of the response of the prototype in high multiplicity showers. In addition, the possibility to distinguish by use of the three thresholds multi-particles and single particle within 1\,cm$^{2}$ cells can be examined. Finally thanks to the possibly large spatial extend of some showers, a uniformity measurement can be carried out quickly.

\begin{figure}[h!]
\begin{centering}
\includegraphics[width=0.3\textwidth]{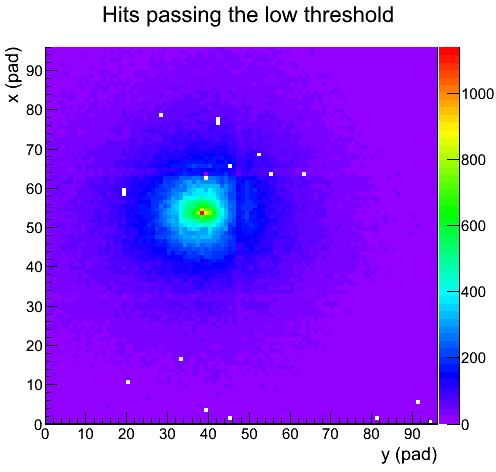}
\includegraphics[width=0.3\textwidth]{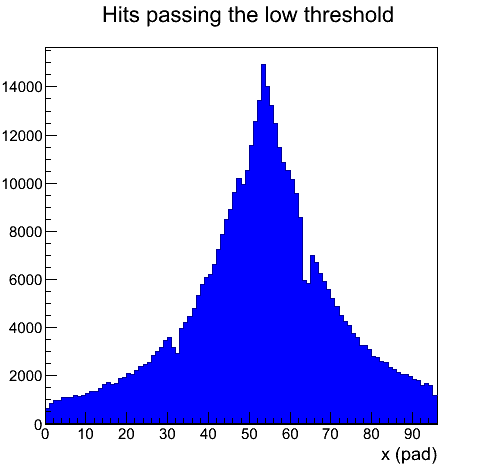}
\includegraphics[width=0.3\textwidth]{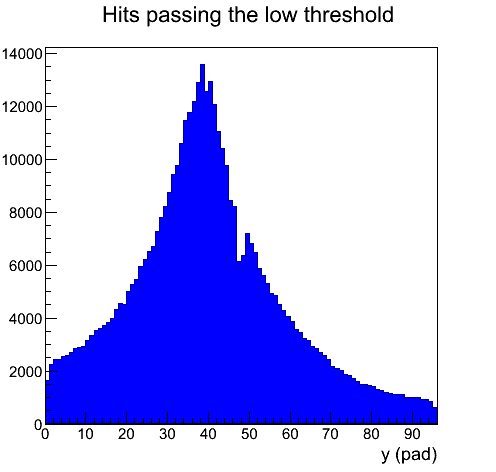}
\caption{Integrated 2D and 1D image of 10$^{4}$ pions of 150\,GeV/c traversing a 1\,$\lambda$ Fe block.}
\label{shower_profile}
\end{centering}
\end{figure}

\paragraph{Uniformity}
The superimposition of 10$^{4}$ pion events measured at a mesh voltage of 350\,V (MIP efficiency of 90\,\%) is depicted in Figure~\ref{shower_profile} (left). The pattern exhibits rings that correspond to non-showering pions in the center surrounded by electromagnetic and hadronic shower components. Thanks to a uniform response, a clear azimuthal symmetry is observed. The effect of dead zones at the ASU junctions is better seen when projecting the pattern along vertical and horizontal axis (Figure~\ref{shower_profile} middle and right). Pads at the ASU edges show a lower efficiency, as expected from the 2\,mm wide Bulk coverlay line running along them.

\begin{figure}[h!]
\begin{centering}
\includegraphics[width=0.3\textwidth]{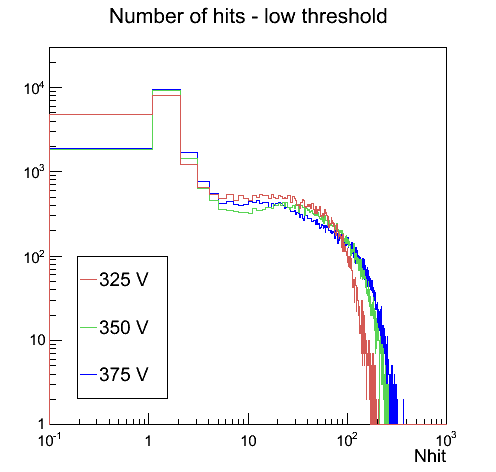}
\includegraphics[width=0.3\textwidth]{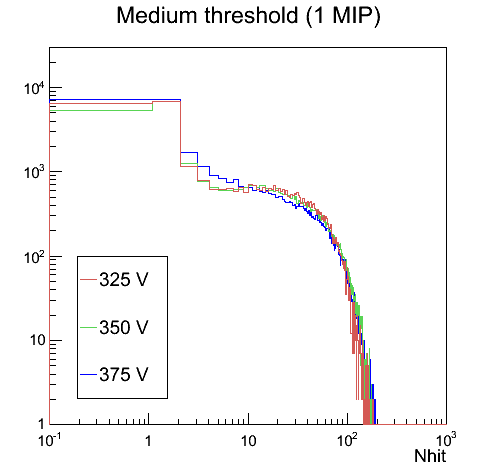}
\includegraphics[width=0.3\textwidth]{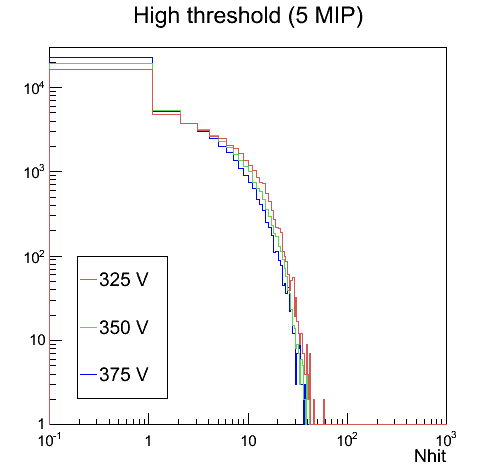}
\caption{Number of hits passing the three thresholds at various voltages.}
\label{shower_hnhit}
\end{centering}
\end{figure}

\paragraph{Hit multiplicity}
\label{mult_h4}
Runs with 5.10$^{4}$ recorded pion events were performed at various mesh voltages.
The primary goal was to measure the hit multiplicity and compare it to Monte Carlo predictions.
Distributions obtained at 325, 350 and 375\,V exhibit a sharp peak and a long tail from traversing and showering pions respectively (Figure~\ref{shower_hnhit} left).
The tail extends to larger values at higher voltages while the population of the zero hit bin drops.
At 375\,V (MIP efficiency of 97\,\%), the maximum number of hits is quite large: up to $\sim$\,300.

\paragraph{Semi-digital readout}
The three thresholds were set to 1\,fC, 1\,MIP and 5\,MIP respectively where the charge of a MIP depends on the mesh voltage. At 325, 350 and 375\,V, a different set of medium and high thresholds had to be used to accommodate for the larger value of the MIP charge at higher voltages. As expected, the number of hits distributions through these two thresholds are very similar, demonstrating a good control of the thresholds.

\subsubsection{ASIC features}

\begin{figure}[h!]
\begin{centering}
\includegraphics[width=0.3\textwidth]{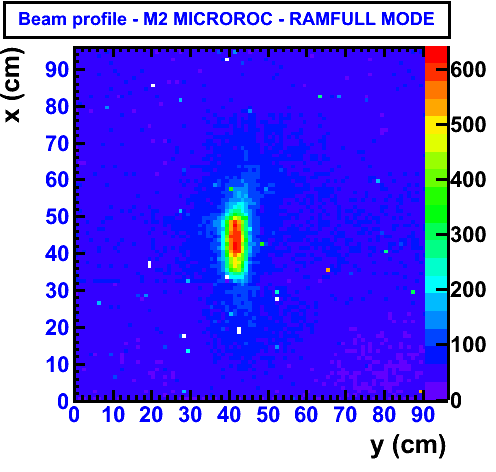}
\includegraphics[width=0.45\textwidth]{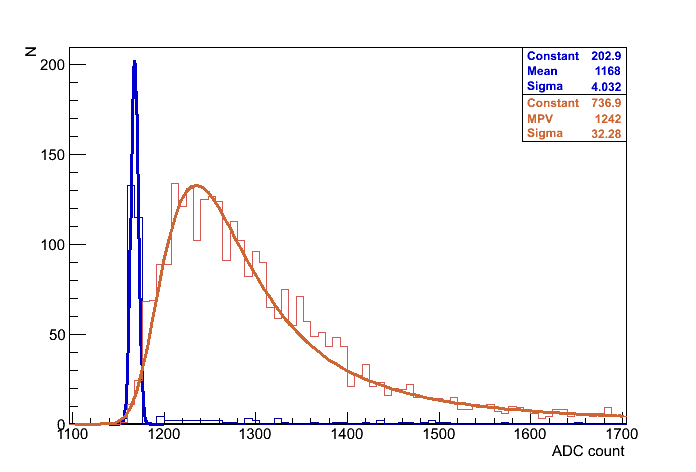}
\caption{Muon beam profile obtained from raw data in trigger-less operation (left). Pion signal distribution measured on one pad with the analogue readout (right).}
\label{ramfull}
\end{centering}
\end{figure}

\paragraph{Trigger-less operation}
When operated in trigger-less mode, the prototype is read out upon receipt at the DIF boards of a memory full signal from one of the 144 ASIC. Stable operation therefore requires negligible noise and discharge rates which could fill memories up and generate dead time. This was easily achieved at 365\,V (MIP efficiency of 95\,\%) when measuring the beam profile (Fig.\,\ref{ramfull}). All information from the detector is used in this plot.

\paragraph{Analogue readout} The possibility to measure the shaper signal with high resolution was implemented with dedicated PCB routing lines and a 12-bit ADC on the DIF boards. Focusing the pion beam on a small region, expected Landau distributions were measured. Figure~\ref{ramfull} (right) shows one distribution with the pedestal peak. This analogue readout is an extra functionality that could be used at ILC/CLIC to monitor pedestals and thresholds. Also, it will be helpful to tune better Monte Carlo simulation parameters.

\section{Conclusion}

Major milestones have been met in 2011 for the validation of the Micromegas technology for hadron calorimetry.
The successful implementation of a new low noise ASIC in a large chamber of 1\,m$^{2}$ size is one of them.
Preliminary test beam results show that this detector provides high sensitivity to MIPs and detailed shower information.
Efforts towards real calorimetry measurements will be pursued with the construction of more planes and the participation to combined CALICE test beams inside m$^{3}$ size Fe or W structures.

\newpage

\section{Bibliography}

\begin{footnotesize}

\end{footnotesize}

\end{document}